# Philosophical perspectives on ad hoc hypotheses and the Higgs mechanism

**Simon Friederich · Robert Harlander · Koray Karaca**

The authors' names are listed in alphabetical order.

S. Friederich
*New address*
University of Groningen, Faculty of Philosophy, Oude Boteringestraat 52,
NL-9712 GL Groningen, The Netherlands
e-mail: email@simonfriederich.eu
*Former address*
Philosophisches Seminar, Universität Göttingen, Humboldtallee 19,
37073 Göttingen, Germany

R. Harlander
Theoretische Teilchenphysik, Fachbereich C, Universität Wuppertal,
42097 Wuppertal, Germany
e-mail: robert.harlander@uni-wuppertal.de

K. Karaca
Interdisciplinary Centre for Science and Technology Studies (IZWT),
University of Wuppertal, Gaußstr. 20, 42119 Wuppertal, Germany
e-mail: karacak@gmail.com

**Abstract:** We examine physicists' charge of ad hocness against the Higgs mechanism in the standard model of elementary particle physics. We argue that even though this charge never rested on a clear-cut and well-entrenched definition of "ad hoc", it is based on conceptual and methodological assumptions and principles that are well-founded elements of the scientific practice of high-energy particle physics. We further evaluate the implications of the recent discovery of a Higgs-like particle at the CERN's Large Hadron Collider for the charge of ad hocness against the Higgs mechanism.

> "Of course our model has too many arbitrary features for these predictions to be taken very seriously [...]" (Weinberg 1967, pp. 1265-1266)

## 1. Introduction

The Higgs mechanism (HM) is a crucial part of the "standard model" (SM) of elementary particle physics. Its main purpose is to account consistently and in accordance with experimental results for the non-vanishing masses of elementary particles in the SM. Recent experimental findings at the Large Hadron Collider (LHC) at CERN in Geneva have been interpreted as a direct detection of a particle with properties as expected for the Higgs particle, which would be a spectacular confirmation of this picture.[1] In its role as part of the SM, involving a single fundamental scalar particle, we refer to the HM in what follows as the "standard model Higgs mechanism" (SMHM).

---

1  On July 4, 2012, ATLAS and CMS, the two largest experimental collaborations at CERN, announced the observation of a new particle whose properties, as measured up to now, agree well with those of a Higgs boson to be Despite its predictive and explanatory success, particle physicists as well as philosophers of physics have expressed qualms about the SMHM. It has been widely regarded as problematic in several important respects. These qualms and worries have often been summarized by calling the SMHM "ad hoc".

The ad hoc charge against the SMHM is interesting from a philosophical point of view on several grounds. First, the claim that our currently best theory of elementary particle physics is based on an ad hoc hypothesis sounds alarming and is certainly worthy of consideration in itself. Second, there exists a longstanding philosophical debate about the notion of ad hocness, to which eminent philosophers of science such as Popper, Lakatos, Schaffner, Grünbaum, Leplin and others have made important contributions. This gives rise to the question of whether the SMHM qualifies as "ad hoc" according to any of these philosophers' accounts of ad hocness, and what the possible ramifications would be. Third, it seems natural to ask what impact the recent experimental discovery of a Higgs-like particle at the LHC has on the status of the ad hoc charge against the SMHM.

In the present work we try to assess this ad hoc charge in a systematic manner. Section 2 gives a brief non-technical introduction to the SMHM. Section 3 reviews physicists' and philosophers' main worries, which underlie the ad hoc charge. Section 4 recapitulates core ideas of philosophical accounts of ad hoc hypotheses put forward by philosophers of science in the past. In Section 5, we show that according to a strict reading of philosophers' accounts, the SMHM does not qualify as ad hoc; however, as we argue further, the criteria which are not obeyed seem to play only a secondary role in how scientists actually classify hypotheses as ad hoc. Disregarding these criteria, Section 6 analyses in detail the SMHM in the light of Jarrett Leplin's account, which is arguably the most refined account found in the literature. We show that it captures nicely the general thrust of the ad hoc charge against the SMHM. In Section 7, we discuss what implications the recent discovery of a Higgs-like scalar particle at the LHC has on the status of the ad hoc charge against the Higgs mechanism. As we argue, this discovery discharges the SMHM from the most straightforward ad hoc allegation, which concerns the previously lacking independent empirical support, but the main criticisms of the SMHM remain valid. On the other hand, as we note, a new tendency in the physics community towards re-evaluating or even discarding these criticisms appears to be visible. Finally, in Section 8, we defend the significance of the concept of ad hocness in evaluating scientific hypotheses against a recent criticism put forward by Christopher Hunt (Hunt 2012).

**2. The Higgs mechanism in a nutshell**

The SM describes the strong and electro-weak interactions among the elementary particles in the language of quantum field theory. It is formulated in terms of a Lagrangian, which has the crucial property of being invariant under certain well-defined transformations of the fields called "local gauge transformations". Local gauge invariance is necessary for the SM Lagrangian to be renormalizable, i.e. mathematically consistent (by the standards of conventional quantum field theory) and predictive up to arbitrarily high energies.

---

expected from the SMHM; see (ATLAS Collaboration 2012) and (CMS Collaboration 2012). Meanwhile, more data have been collected and analyzed, and the "observation" has been upgraded to a "discovery"; see (ATLAS Collaboration 2013) and (CMS Collaboration 2013).

The SM employs the HM to account for non-zero particle masses in the Lagrangian, which would otherwise violate gauge invariance and thus spoil the renormalizability of the theory. Its basic idea is the existence of a scalar field (i.e. a field with spin 0), the so-called Higgs-field, which has an infinite number of degenerate lowest-energy field configurations. Classically, this field has a non-zero vacuum expectation value, in which respect it differs from all other fields of the SM Lagrangian. This is often expressed by saying that local gauge invariance, though not violated explicitly, is absent on the level of the ground-state of the theory in the form of a so-called "spontaneous breaking" of the local gauge symmetry.[2]

A consequence of the features just described is that particle masses are induced through the interaction between the particles and the Higgs field. The SMHM thereby makes the prediction that the strength of a particle's interaction with the Higgs boson is proportional to its mass. As a consequence, for example, the ratio of the Higgs decay rate into particles such as muons and bottom-quarks is equal to the (square of the) ratio of the muon and bottom-quark masses, up to calculable effects due to kinematics or higher order quantum fluctuations.

There is quite an amount of freedom as regards how the HM is implemented in the electro-weak theory (and thus in the SM). In particular, the behavior of the Higgs field under gauge transformations, its so-called gauge representation, is not fixed by theoretical requirements. Steven Weinberg was the first to use the HM in order to preserve gauge symmetry in electro-weak interactions. In his seminal paper (Weinberg 1967), usually considered the birth of the SM, he implemented the HM in its minimal form by including a single Higgs field in the simplest possible gauge representation. This is what we refer to as the SMHM in this paper. Other, more complicated, representations were considered later in history, but they either require a larger number of parameters (e.g. two-Higgs doublets), or are in conflict with precision measurements. We will therefore focus on the SMHM, although most of our discussion applies to other implementations as well. In particular, the aspects of the ad hoc charge discussed in Section 3 all refer to its field-theoretic aspect, not to the particular gauge representation.

Quite often, probability (or unitarity) conservation is referred to as another benefit of the SMHM, besides renormalizability. Unitarity conservation means the following: For vector particles like the W and Z bosons, non-zero masses are equivalent to the existence of longitudinal degrees of polarization. If gauge invariance is broken on the level of the Lagrangian, these components can contribute to scattering amplitudes that "violate unitarity", meaning that the probability of the respective scattering reaction becomes ill-defined. If these masses are induced by the SMHM, however, the Higgs particle's contribution to the scattering amplitude exactly cancels the unitarity-violating terms. Both renormalizability and unitarity conservation are consequences of the same feature of the Lagrangian, namely, gauge invariance.

Interactions among the particles lead to observable processes involving the Higgs particles, which are detectable at particle colliders. The predicted phenomenology of effects due to the Higgs particles has been studied in great detail in the physics literature (see, e.g. (LHC 2011)), and the experimental search program has been active for several decades. Indirect searches in the 1990s based on the comparison of precise measurements made by the LEP experiment at CERN and the SLD experiment at SLAC with quantum field theoretical calculations provided first clues for the value of the Higgs boson mass (see, e.g. (LEP 2003)). Direct searches were subsequently done at the LEP2 experiment with increased beam energy, leading to a lower Higgs mass limit of about 114 GeV. Further mass exclusion limits have been obtained since 2008 at the Tevatron collider at Fermilab, and later at the LHC at CERN. As mentioned already above, on July 4, 2012, a signal for a particle at around 125 GeV was observed. Its properties, as measured so far, are compatible with a SM Higgs boson.

---

[2] This characterization of the SMHM in terms of spontaneous symmetry breaking and a degenerate ground state has to be taken with a grain of salt and is in some respects misleading; see (Smeenk 2006), (Healey 2007, Chapter 6.5), and (Friederich 2013) for clarifications aimed at philosophers.

**3. Physicists' criticisms and accusations of ad hocness against the SMHM**

Even though the SMHM forms part of the SM as a renormalizable quantum field theory, which, to the current date, describes all phenomena observed at particle colliders to very high precision, physicists have searched for alternatives to a fundamental Higgs field already early on. Evidently, until very recently one of the main motivations for this was the lack of direct experimental evidence for the Higgs boson. However, even now after the recent discovery of a Higgs-like particle, many particle physicists would consider the confirmation of this particle being the Higgs boson with properties exactly as predicted by the SM as a disappointment. In fact, the situation that the LHC discovers a SM Higgs boson and nothing else is often called the "nightmare scenario" by many physicists (see Cho (2007)).

We begin by citing an objection against the SMHM that was mostly relevant prior to the recent discovery:

1. *There is a lack of direct evidence for the SMHM.*

Before the recent detection event at the LHC, this criticism was very natural: Among all the different types of particles in the SM, the Higgs particle is the only scalar particle and it has a unique status and role among all SM-particles. At the same time, it was the only type of particle for which no direct evidence existed. In the words of Lee Smolin, this endowed the SMHM with an "ad hoc quality":

> [An] obstacle arises from the [SM's] reliance on the idea of spontaneous symmetry breaking to explain why each of the elementary particles we see in the world has different properties. While it is a beautiful idea, there is a certain ad hoc quality to how [the SMHM] is realized. To this date, no one has so far observed a Higgs particle and we have only a very imprecise idea of their actual properties. (Smolin 1997, p. 61)

We will later (Section 8 in particular) discuss what impact the recent discovery has on this criticism. Smolin's complaint concerning how spontaneous symmetry breaking "is realized" in the SMHM may refer not only to the lack of direct evidence for the Higgs particle but also to the fact that the Higgs particle is a *fundamental* scalar particle and that symmetry breaking in the SMHM is not of *dynamical* origin. Both points are discussed in what follows, one immediately as criticism 2, the other further below as criticism 4.

2. *There are no other fundamental scalar particles besides the Higgs boson in the SM, nor is there any experimental evidence for such particles.*

Except for the Higgs boson, all particles of the SM – and thus all known fundamental particles – either have spin ½ (quarks and leptons) or spin 1 (gauge bosons: W, Z, gluons, and photon). Lorentz invariance as required by special relativity, on the other hand, demands that the Higgs field be scalar, i.e. have spin 0. All known scalar particles are composed of more fundamental ones; examples are pions and kaons, which are composed of quarks and gluons. Concern about the SMHM on these grounds is expressed, for example, by the MIT physicists Edward Farhi and Roman Jackiw, who criticize the introduction of the Higgs field as a fundamental scalar field as "ad hoc":

> While the Weinberg-Salam model [i.e. the SM] is recognized to be a theoretical and experimental success, it is frequently believed that the Higgs mechanism [...] is an unsatisfactory feature of the theory. There is no experimental evidence for fundamental scalar fields, which are introduced in an *ad hoc* manner with *ad hoc* interactions solely to effect the symmetry breakdown. There is no other compelling theoretical reason for scalar fields [...] (Farhi and Jackiw 1982, p. 1)

In a similar vein, Andrei A. Slavnov notes that "[the SMHM] is based on the ad hoc introduction of scalar fields which appear as fundamental elementary particles together with leptons, quarks and Yang-Mills mesons [i.e. the particles mediating the interactions between quarks and leptons]", and he adds that this introduction of fundamental scalars is not an "attractive possibility."(Slavnov 1979, 289) Two widely shared worries about the introduction of fundamental scalars to which Slavnov seems to allude to here are discussed further below, namely, the so-called naturalness and triviality problems.

3. *The conception of the vacuum as pervaded by a non-vanishing Higgs field is conceptually problematic.*

As explained in the previous section, the SMHM is often characterized by saying that it involves a ground state degeneracy and leads to a non-zero expectation value for the Higgs field. This aspect of the SMHM is sometimes criticized as conceptually suspect (though not explicitly "ad hoc", as far as we know). For example, in an introductory textbook on gauge theories physicist K. Moriyasu compares the "Higgs field [...] to an old-fashioned 'aether' which pervades all space-time [and] acts like a continuous background medium even at very short distances." (Moriyasu 1983, p. 120)

In a similar vein, philosopher of science Margaret Morrison alludes to the assumption that the Higgs field has a non-zero vacuum expectation value and contends that here "we are dealing with fields whose *average value* is non-zero, where the vacuum is said to have non-zero expectation value." (Morrison 2003, p. 359) Highlighting this aspect of the SMHM, she characterizes the vacuum of the SM as a "plenum" (Morrison 2003, p. 357) and argues that "the various vacuum hypotheses which provide the necessary theoretical foundation [of the SMHM] are essentially problematic, for both physical and philosophical reasons." (Morrison 2003, p. 361)

However, independently of why Morrison deems "the various vacuum hypotheses [...] problematic", her complaint about the degenerate vacuum and the associated non-zero vacuum expectation value is misplaced, unless it is understood metaphorically as a mere rephrasing of the other criticisms mentioned. We do not want to enter into the details of a discussion about gauge symmetry breaking and the claimed vacuum degeneracy. But, in order to indicate that matters are more subtle than suggested by Moriyasu and Morrison, it suffices to note that the Higgs field itself is not gauge invariant, which means that its vacuum expectation value is gauge-dependent and thus not a physical quantity. Consequently, whether or not that vacuum expectation value is non-zero under certain conditions does not by itself have any physical import. This already indicates that the SMHM does not substantially rely on attributing a non-zero vacuum expectation value to the Higgs field and thus not on a questionable "theoretical story about the nature of the vacuum" (Morrison 2003, p. 361), contrary to what Morrison – and similarly Moriyasu – seem to suggest.[3]

4. *All other known cases of symmetry breaking in nature are dynamical, that is, due to composite, rather than fundamental fields. In the SMHM, in contrast, a dynamical account is lacking and the symmetry breaking is implemented by fiat through the shape of the Higgs potential.*

The conceptual framework of the HM is of theoretical use not only in particle physics, but also in condensed matter physics, where, for example, the HM functions as a conceptual tool to account for phenomenological features of superconductors. In that case, the field that plays the role of the Higgs field is not fundamental, but associated with pairs of electrons (so-called Cooper pairs), which are held together by the complex interplay of electro-magnetic interactions. Within particle physics, a similar situation is found in chiral symmetry breaking in QCD due to what is called a "quark condensate", which is a composite system just like the system of the Cooper pairs. Symmetry breaking which occurs due to a field associated with composite particles is said to be *dynamical*. In the SMHM, in contrast, where the Higgs field is fundamental rather than composite, the symmetry breakdown due to the Higgs field is non-dynamical. According to the following passage by Jackiw, this feature of the SMHM makes it ad hoc:

> Spontaneous symmetry breaking is adopted from many-body, condensed matter physics, where it is well understood: the dynamical basis for the instability of symmetric configurations can be derived from first principles. In the particle physics application, we have not found the dynamical reason for the instability. Rather, we have postulated that additional fields exist, which are destabilizing and accomplish the symmetry breaking. But this *ad hoc* extension introduces additional, *a priori* unknown parameters and yet-unseen particles, the Higgs mesons [i.e. Higgs bosons]. (Jackiw 1998, p. 12777)

As emphasized by Jackiw, unlike in condensed matter physics, in the SMHM symmetry breaking is implemented non-dynamically through the choice of parameters in the so-called "Higgs potential" (the part of the Lagrangian which accounts for the mass of the Higgs boson and its self-interaction). For this reason, physicist Bruce Schumm characterizes this potential as an "arcane and ad hoc notion" in itself (Schumm 2004, p. 329). According to theoretical physicist Michael Peskin, since the parameters of this potential are put in by hand, "we should be ashamed of ourselves if we are satisfied" (Peskin 2012, p. 12) with the understanding of mass generation provided by the SMHM.

5. *The SMHM leads to a large number of independent parameters for the SM, and it does not explain their values or reduce their number.*

---

3  These considerations are enforced by noting that from a gauge-independent perspective, local gauge symmetry is never broken and the Higgs field, as a gauge-dependent variable, has zero expectation value under all conditions. See (Elitzur 1975) for an important result on this matter and (Friederich 2013, Sects. 5 and 6) for a more detailed discussion of the ramifications of this and other results that is aimed at philosophers.

Not only the parameters of the Higgs potential but also the couplings between the Higgs particle and the other particles of the SM (which are directly related to their masses, as explained in Section 2) cannot be derived from more fundamental principles. They must all be introduced as independent parameters on the basis of experimental data without further explanation for their values.[4] According to CERN theorist Gian Francesco Giudice, this is highly unsatisfactory (where the last sentence in the quote reiterates the previously discussed criticism concerning the non-dynamical character of spontaneous symmetry breaking in the SMHM):

> Unlike the rest of the theory, the Higgs sector is rather arbitrary, and its form is not dictated by any deep fundamental principle. For this reason its structure looks frightfully ad hoc. [...]
>
> The Higgs sector explains the structure of quark and lepton masses that we observe, but only at the price of introducing 13 adjustable input parameters determined by experimental measurements. Quark and lepton masses can certainly be accounted for by the Higgs sector, but unfortunately the theory is unable to predict their values. Moreover, although the Higgs sector can generate the spontaneous breaking of electroweak symmetry, it provides no deep explanation about the force that is ultimately responsible for the phenomenon. (Giudice 2010, p. 174)

It should be mentioned, however, that the main alternatives to the SMHM enhance the number of independent parameters even further, and they do not explain any of the curious structures just mentioned. The minimal supersymmetric extension of the SM, for example, which is probably the most studied model beyond the SM, requires more than 100 new parameters. So, the SMHM is in fact less problematic than its most-discussed alternatives as far as the criticism raised by Giudice is concerned. Most of these alternatives are conceived as responses to the criticism of the SMHM to be discussed next, the so-called "naturalness problem".

6. *Fine Tuning and Naturalness*

The argument against the fundamental scalar character of the Higgs field that is widely regarded as the most severe is the so-called "naturalness" or "fine tuning" problem (FTP). Its precise formulation is within the renormalization formalism and thus beyond the scope of this paper (see (Susskind 1979) for a classic reference). We will nevertheless try to give a qualitative description: According to quantum field theory, any particle mass measured in experiment (the so-called physical mass) can be seen as the sum of the bare mass and a contribution due to the interaction of the particle with so-called vacuum fluctuations, called its interaction mass. Usually, one assumes the theory to be valid only up to a (large but finite) energy scale, the "cut-off" (which can simply be the mass of a particle not predicted by the SM), beyond which effects accounted for by a more fundamental (yet presently unknown) theory are supposed to set in. One can calculate the interaction mass as a function of that cut-off. The bare mass then follows as the difference between the physical and the interaction mass.

---

[4] Even though the values of parameters used in the SMHM are theoretically unconstrained and therefore arbitrary, they appear by no means random. For example, the so-called CKM matrix which relates the three quark generations in the SM, is approximately diagonal, even though its elements are not in any way determined by the theory and could therefore be arbitrary. Other oddities are the enormous spread of the charged-fermion masses, spanning six orders of magnitude, with the largest one (the top-quark mass) being suspiciously close to the Higgs field's vacuum expectation value.

For spin ½ and spin 1 particles, the dependence on the cut-off is logarithmic, leading to an interaction mass that is at most of the same order of magnitude as the bare mass. Bare and physical masses are thus typically also of the same order of magnitude. For scalar particles, on the other hand, the dependence of the interaction mass on the cut-off is quadratic. Assuming the SM to be valid up to the Planck energy scale ($10^{19}$ GeV), the bare mass and the effect of the vacuum fluctuations would have to balance each other by about 34 orders of magnitude in order to result in a physical Higgs mass of 125 GeV. The disturbing fact about such a large cancellation is that the bare mass of the Higgs is a free parameter; the reason why nature would have fine-tuned it to almost exactly the interaction mass is unexplained by the SMHM. As a consequence, the SM appears "unnatural" in this respect, which is why the problem is called the "naturalness problem".

The FTP arises from the scalar nature of the Higgs field and is therefore directly linked to the ad hoc charge against the introduction of a scalar Higgs field. It is unsurprising that the most popular models devoid of the FTP solve at least one of the issues outlined before as well: they assume either a non-fundamental, composite Higgs boson and provide a dynamic account of gauge symmetry breaking, or they try to explain the shape of the Higgs potential (e.g. in the case of supersymmetry; the term which is quadratic in the Higgs field is driven negative by its renormalization group evolution). However, none of them seems to be presently privileged with respect to the SM by available experimental data, and many of them are ruled out.

7. *Triviality*

Finally, let us mention a further issue related to the Higgs boson, which is also considered unsatisfactory by many physicists. Again, its precise formulation requires the renormalization group formalism, but even then the argument requires approximations and extrapolations (see (Callaway 1988)). The worry is that the self-coupling of the Higgs field may diverge at large but finite momentum transfers, which would mean that the SM cannot be consistently extrapolated to arbitrary high energy scales and breaks down at or below the corresponding energy. The precise position of this so-called "Landau pole" depends on the Higgs mass. For the experimentally observed mass $M_H \approx$ 125 GeV, the Landau pole is completely absent and therefore does not put a limitation on the SM. The name "triviality" reflects the fact that in pure scalar theory, i.e. one which contains only the Higgs field and no other particles, the Landau pole can only be made to disappear completely if the self-coupling of the Higgs boson is assumed to be vanishing, i.e. trivial, while this self-coupling must be non-vanishing for the HM to generate non-vanishing particle masses.

**4. Philosophical accounts of ad hoc hypotheses**

The notion of an ad hoc hypothesis[5] owes its widespread use to Karl Popper ((Popper 1959), Sections. 5, 19, and 46) who employed it to characterize hypotheses which are added to scientific theories in response to nonconforming empirical data. According to him, "a conjecture [is] "*ad hoc*" if it is introduced […] to explain a particular difficulty, but if […] *it cannot be tested independently*" (Popper 1974, p. 986), thus reducing the degree of falsifiability of the theory.

---

5   See (Karaca 2010), Chapter 6, for a more detailed overview of philosophical accounts of ad hoc hypotheses.

Kenneth Schaffner departs from Popper in that he calls not only those hypotheses "ad hoc" which *cannot* be tested independently, but also those for which, relative to a given scientific context, there is no "independent theoretical or experimental support." (Schaffner 1974, p. 68) Thus, for him, "*ad hocness* [is] a property of specific hypotheses, as embedded [...] in a constellation of other hypotheses constituting the theory" (Schaffner 1974, p. 67), not a feature of hypotheses in themselves. Adolf Grünbaum (1976) also conceives of ad hocness as context-relative by taking the ad hoc character of an ad hoc hypothesis to be time-dependent. On his account, an ad hoc hypothesis at a time t may no longer qualify as ad hoc at a later time t* – provided that the theory which incorporates it makes the right kind of empirical progress between the two times. We shall later consider in which sense this may hold for the SMHM.

The account of ad hoc hypotheses which we choose as the basis of our discussion of the SMHM is due to Jarrett Leplin (1975). In what follows we review the criteria which he proposes as separately necessary and jointly sufficient for auxiliary hypotheses to qualify as ad hoc.

Leplin's first condition is familiar from the accounts of Popper, Schaffner, and Grünbaum:

> *Condition of experimental anomaly*
> If an hypothesis H is introduced into a theory T in response to an experimental result E, then if H is *ad hoc*, E is anomalous for T but not for T as supplemented by H. (Leplin 1975, p. 317)

In Leplin's sense, the phrase "in response to" is to be understood as implying that T was known before E. This excludes situations where the suggested theory T is in conflict with existing experimental data E and gets "rescued" from this conflict by an auxiliary hypothesis H at some point. We will see that already this restriction (henceforth referred to as "T before E") prevents the SMHM from being ad hoc according to Leplin.

Note also that the condition of experimental anomaly does not require E to be the only aspect where T is in disagreement with experiment. This will be used explicitly in the condition of non-fundamentality further below.

Let us remark on a certain lack of definition in the expression "an experimental result": it is unclear whether this includes only a single number (say, the maximal brightness of a certain type of supernovae), a distribution of numbers (e.g., the time evolution of the brightness of these supernovae), or maybe even more general data samples (the time evolution of all types of supernovae). The fuzziness of this condition affects its application to the SMHM, as we will see later.

Also the second of Leplin's conditions is in agreement with the previously stated accounts of ad hoc hypotheses. It concerns their context of justification, exemplified in various historical cases of theory modification, including, e.g., the Lorentz-Fitzgerald contraction hypothesis, Pauli's neutrino hypothesis and the hypothesis of a trans-Uranian planet by Adams and Leverrier. In Leplin's words, an "*ad hoc* hypothesis is one introduced in response to an experiment that provides its only support" (Leplin 1975, p. 319).[6] We consider this condition the central condition of ad hocness, carrying the actual meaning of the Latin expression "ad hoc" (="for this"):

> *Condition of justification*
> If an hypothesis H is introduced into a theory T in response to an experimental result E, then if H is *ad hoc*, E is evidence for H but:
> 1. No available experimental results other than E are evidence for H.
> 2. H has no application to the domain of T apart from E.
>
> 3. H has no independent theoretical support. (Ibid., p. 320)

Here, what Leplin means by "the domain of a theory" is "a set of propositions specifying such requirements for T: reports of experimental results, statements of problems, hypotheses of previous theories, descriptions of postulated entities" (Ibid., p. 318).

In Leplin's view, an ad hoc hypothesis is added into a scientific theory without destroying its internal coherence. This results in the following

> *Consistency condition*
> If an hypothesis H introduced into a theory T is *ad hoc*, then H is consistent with accepted theory and with the essential propositions of T. (Ibid., p. 327)

Here, Leplin defines an "essential proposition" of a theory to be a proposition whose rejection would count, in the judgment of the scientific community, as a rejection of T irrespective of the retention of other hypotheses (Ibid., p. 327).

Leplin does not think that the features of ad hoc hypotheses presented in the previous conditions are by themselves sufficient for endorsing or rejecting those hypotheses. What he calls "condition of tentativeness" states that the allegation of ad hocness against a hypothesis should not be construed as a commitment to either its future confirmation or disconfirmation:

> *Condition of tentativeness*
> If an hypothesis H is *ad hoc*, then there are no sufficient grounds for holding that H is true and no sufficient grounds for holding that H is false. (Ibid., p. 321)

This condition is the hardest one to test in general, since it is usually very difficult to decide in which cases there actually are sufficient grounds for holding that a hypothesis is true. However, we will see that for the SMHM the condition of tentativeness can be construed as met in a rather well-defined and broadly accepted way.

Unlike the previous conditions, the last condition Leplin sets forth concerns the way in which an experimental anomaly is understood prior to its treatment by an ad hoc auxiliary hypothesis:

> *Condition of non-fundamentality*

---

6  The fact that the neutrino and the trans-Uranian planet hypotheses were *independently* "confirmed" in the past and, as a result of this, no charge of "ad hocness" is raised against them any more seems to vindicate the general consensus on ad hoc hypotheses indicated by Leplin's *condition of justification*.

> If an hypothesis H introduced into a theory T in response to an experimental result E is *ad hoc*, then there are problems other than E confronting T which there is good reason to believe are connected with E in the following respects:
>> 1. These problems together with E indicate that T is non-fundamental.
>> 2. None of these problems including E can be satisfactorily solved unless this non-fundamentality is removed.
>> 3. A satisfactory solution to any one of these problems will contribute to the solution of the others. (Ibid., p. 331)

Again, this condition carries a significant amount of arbitrariness, manifest in the expressions "good reason", "believe", and "satisfactory". Also, it is left unspecified how severe the "problems other than E" need to be.

Lastly, we stress that Leplin restricts the use of "ad hoc hypothesis" to cases in which a hypothesis criticized as "ad hoc" is introduced as a response to a disconfirming experimental result. Leplin is aware that this requirement limits his analysis to cases of "experimental" rather than "theoretical anomaly". However, he does not think that this limitation takes away from the generality of his analysis. Leplin's consideration here is that even though in principle there could be cases in which some hypothesis is properly classified as "ad hoc" without being connected to any experimental anomaly, no such case has yet arisen in scientific practice. However, as we shall argue later, the SMHM constitutes precisely such a case.

**5. The SMHM is not strictly speaking ad hoc according to Leplin's account**

It is clear that the SMHM cannot be ad hoc according to accounts, such as Popper's, that suggest that ad hoc hypotheses are in principle unable to make independently testable predictions. The SMHM made at least one such prediction (which turned out to be *actually* testable), namely, that a neutral scalar particle with spin 0 should exist. Therefore, according to any account which rules out that ad hoc hypotheses may make independently testable predictions, the SMHM would not count as ad hoc. However, it is far less obvious (and much more rewarding to investigate) how the SMHM fares according to more liberal accounts of ad hocness such as Leplin's.

Let us introduce the notation "SM0" for the SM without the Higgs sector, implying the symbolic equation SM=SM0+SMHM. The SM0 is a well-defined theory no less than the SM itself, fulfilling all the properties of a consistent quantum field theory, such as renormalizability, absence of anomalies, and unitarity. However, the kind of physical world that it describes is radically different from our own: In particular, all the particles it accounts for are massless. In addition, a direct consequence of the SM0 is that the notion of electric charge is no longer physical; rather the confining character of the non-Abelian weak interaction would make it possible to physically distinguish the weak charge from the hypercharge. In contrast to electric charge, hypercharge is the same for the electron and the neutrinos, for example, which would be indistinguishable in the SM0.[7]

Certainly, the SM0 has never been proposed as a candidate theory of elementary particle physics for our world. This is the first reason why, strictly speaking, the SMHM does not qualify as an ad hoc hypothesis according to Leplin. Contrary to the criteria he proposes, physicists do not seem to regard it as obligatory for the hypothesis H to qualify as ad hoc that the "pre-ad hoc" theory T has ever been held as a serious candidate theory in its own right; it seems fully sufficient that T be, in some hardly specifiable sense, a viable theory.

---

7  In Section 5, "A world without the Higgs mechanism", of (Quigg 2007), Chris Quigg offers more detailed considerations as to what a world described by the SM0 would be like.

On the other hand, in a slightly looser sense, we can in fact find an earlier theory that may be considered as the "pre-ad hoc" version of the (electroweak part of the) SM, namely Glashow's model[8] (Glashow 1961), which is based on the same gauge group SU(2) x U(1). Disregarding the mass terms which Glashow introduced at the cost of breaking gauge invariance explicitly (rather than spontaneously), it results in something very close to the SM0. In that sense, the SM0 "existed" before the SM, whose origin is usually taken to be Weinberg's paper of 1967.

Nevertheless, again taking Leplin's conditions at face value, the SMHM does not qualify as an ad hoc hypothesis added to the SM0 for a second reason. As discussed above, Leplin's conditions for an ad hoc hypothesis involve the restriction "T before E": an existing theory runs into conflict with "new" experimental data and is subsequently "rescued" by an auxiliary hypothesis. However, it was clearly known long before Glashow's paper that the electron mass is non-zero, and also that electron and neutrino are distinct (both of which could be taken as the "experimental result E"). After all, this is why Glashow did not actually suggest the SM0, but introduced explicit mass terms. Therefore, considering the SMHM, also the requirement made by Leplin that H be introduced into an existing theory T in order to count as (possibly) ad hoc, does not conform to how physicists actually characterize hypotheses as "ad hoc".[9]

In a final attempt to argue that Leplin's criteria are literally met by the SMHM, one may argue that Weinberg's theory was an ad hoc modification not of the SM0, but of Glashow's actual theory, i.e. a theory which includes explicitly gauge symmetry-breaking mass terms.[10] Eliminating these mass terms and instead accounting for particle masses by the SMHM renders the theory renormalizable (as later shown by 't Hooft (1971)) and, as such, mathematically consistent and predictive up to arbitrarily high energies.[11] However, Leplin's conditions do not include such a situation where an auxiliary hypothesis cures an internal, or mathematical, deficit of a theory; the hypothesis is always assumed to remove a disagreement with experimental data. This is the third reason why the SMHM does not qualify as ad hoc on Leplin's account.

To summarize the discussion up to this point: No matter whether we consider the SM as a modification of (a) the SM0 or of (b) Glashow's theory, the constraint "T before E" prevents the SMHM from being straightforwardly ad hoc by Leplin's *condition of experimental anomaly*. The same condition gives rise to additional objections against the ad hoc charge for option (a) from the requirement that H be introduced into a previously established theory, and for option (b) from the fact that the SMHM solves a theoretical, rather than an experimental, difficulty.

---

8  In 1979 S. Glashow shared the Nobel prize with S. Weinberg and A. Salam "for their contributions to the theory of the unified weak and electromagnetic interactions between elementary particles" (http://www.nobelprize.org/nobel_prizes/physics/laureates/1979/)

9  See (Karaca 2010), Chapter 7, for the original statement of the claim that the SMHM is not strictly speaking ad hoc on any account according to which ad hocness requires that a conflict be removed between an established theory and incoming empirical data. Rather, Karaca argues that the SMHM is an ad hoc hypothesis that was proposed to solve a conceptual problem, namely, the problem of how to account for the non-zero masses of vector bosons while preserving the gauge invariance of the theory.

10  Whether Weinberg himself considered his theory a modification of Glashow's seems difficult to decide. His only reference to Glashow's model is in a footnote, where he characterizes it as "similar to [his own model]; the chief difference is that Glashow introduces symmetry-breaking terms into the Lagrangian, and therefore gets less definite predictions." (Weinberg 1967, p. 1266)

11  The failure of renormalizability in Glashow's theory may have been the chief reason why it was not recognized earlier as an important contribution. According to the Science Citation Index (Glashow 1961) was cited only once per year between 1961 and 1967. In fact, even Weinberg's paper was cited less than five times in the three years between its publication and 't Hooft's proof of renormalizability ('t Hooft 1971). For a historical comparison of Glashow's and Weinberg's theories, see (Karaca 2013).

Apparently, from the perspective of many physicists, all of these reasons one might have for not regarding the SMHM as ad hoc are only of circumstantial importance.

Instead of proposing an explicit modification of Leplin's criteria, in particular the *condition of experimental anomaly* in response, we will simply *assume* that it can be formulated in a slightly more permissive way. Its precise formulation will require a thorough study of all its implications, as well as a comprehensive confrontation with historical examples. Such an analysis is certainly beyond the scope of the current paper and would divert its focus from the actual subject. Suffice it to say that the refined version should somehow capture the basic idea that, in order for a hypothesis H to qualify as ad hoc, there has to be some (conceptually) viable theory T which suffers from a certain shortcoming E (not necessarily arising from experimental data), while T supplemented by H does not. This idea evidently holds true for the SMHM (with either T=SM0 or T=Glashow's theory[12] and H=SMHM). In the next section, we will see the relation of the SMHM with respect to what we regard as the more essential among Leplin's criteria.

**6. In which sense the Higgs mechanism is ad hoc**

In this section, we evaluate the SMHM with respect to each of Leplin's criteria of ad hocness, except for the condition of experimental anomaly, which we have already discussed in the previous section. It may be helpful for the reader to explicitly recall the individual conditions as we go through them one-by-one in what follows.

*Condition of consistency*

Obviously, Leplin's condition of consistency is met by the SM. The crucial significance of the SMHM can be seen in the fact that it allows to formulate a renormalizable gauge theory that accounts for non-vanishing particle masses. Its most important achievement is to make the SM mathematically consistent and predictive up to arbitrarily high energies.

*Condition of tentativeness*

One way to see the tentative character of the SMHM is as follows: A general quantum field theory involves non-renormalizable field operators and therefore has a built-in restriction of its range of validity. In this case, there clearly are "sufficient grounds" for holding that the theory is, strictly speaking, "false", namely, that it is at best "effective" in the sense of being a mere approximation to a more complete theory with a larger range of validity. In contrast, the SMHM does not involve non-renormalizable operators, so this argument does not apply. However, this does not provide sufficient grounds for holding that it is true, either; for all we know, the SM, as based on the SMHM, might still be only an effective theory and break down at energies far below the Planck scale.

---

12 To arrive at the SM from the starting point of the SM0 is conceptually more straightforward than to arrive at the SM from the starting point of Glashow's theory, which, however, is closer to the actual historical course of events. Thus, whether one prefers to conceive of the SM as an ad hoc modification of Glashow's theory or of the SM0 depends, respectively, on whether one prefers a historical perspective or one of rational reconstruction. (We would like to thank an anonymous referee for proposing this distinction.)

Note that most physicists would agree that the tentative character of the SMHM cannot be decoupled from that of the rest of the SM; if it turns out that the SMHM is in fact only an effective description, a more fundamental mechanism would probably affect all of the SM. In the case of the SMHM, the tentativeness condition is therefore tightly connected to the condition of non-fundamentality, which we discuss next.

*Condition of non-fundamentality*

The SMHM is very widely suspected to be non-fundamental among physicists, namely, in the words of Ross and Veltman, as "something like an effective description of a much more complicated situation" (Ross and Veltman 1975, p. 136). A theory that accounts for the supposed "more complicated situation" could be one with dynamical symmetry breaking, for example, or a supersymmetric theory. In that sense the non-fundamentality charge is reflected in the criticism number 4 in Section 3.

However, also the criticism number 5 in Section 3 falls into this category. The SM suffers from a number of explanatory "problems" related to the number and distribution of its parameters, in particular those associated with the particle masses. Unless one contents oneself with accepting that these parameters are completely random and accidental, they indicate that the SM is non-fundamental. Criterion number 1 of the condition of non-fundamentality is therefore met. By construction, the SM has no explanation for these "problems", which corresponds to criterion number 2. And since many of these problems are directly related to the Higgs sector of the SM, also criterion number 3 is met.

Arguably, also the naturalness (and potentially the triviality) problem suggests that the SM is non-fundamental. Both are linked to the scalar nature of the Higgs particle. So, if one could solve the mass problem of the SM without invoking scalar particles, one would most likely avoid all the criticisms of the SMHM listed in Section 3. This is exactly the spirit of criterion number 3 of Leplin's condition of non-fundamentality.

*Condition of justification*

As already mentioned in Section 4, the condition of justification is a key condition for classifying auxiliary hypotheses as ad hoc. Its wording, however, leaves a lot of room for interpretation. Like the condition of non-fundamentality, it involves three criteria. Let us start with the second and third criteria, before we discuss the first one, which gives rise to the most interesting considerations in our case.

Criterion number 2 (that H should have "no application to the domain of T apart from E") is perfectly reflected in our criticism number 4 of the SMHM: Non-dynamical spontaneous symmetry breaking is used in particle physics only to account for the non-vanishing particle masses through the SMHM; it has "no application" apart from that. One may argue that an additional application of the SMHM is to save the SM from unitarity violation at high energies. However, this problem is so tightly connected to gauge invariance that it cannot be seen as independent; the SMHM preserves gauge invariance, and, therefore, unitarity is guaranteed.

Criterion number 3 (that H "has no independent theoretical support") matches criticism number 2 of Section 3: The Higgs particle has a peculiar status as the only fundamental scalar particle among all the particles of the SM. This can be naturally paraphrased by saying that the hypothesis that there are fundamental scalars has no "independent theoretical support" other than being able to account for non-vanishing particle masses.

Let us now turn to criterion number 1 (that "[n]o available experimental results other than E are evidence for H"). It is here where the recent discovery has crucial implications for the ad hoc charge against the SMHM. We will first take a pre-discovery point of view and discuss the impact of the discovery afterwards.

Criterion number 1 refers to some "experimental result E" which in the case of the SMHM can be taken as the observation of non-zero particle (in particular, weak gauge boson) masses (recall that we decided to ignore the fact that this observation had been made long before the SM was formulated). This choice is not unique, however; one could also argue for the distinguishability of the electron and the neutrino, or other phenomena related to the Higgs sector of the SM. The question is now whether the fact that there actually are several phenomena connected to the Higgs sector is in contradiction with the first criterion of the condition of justification.

The most crucial of these phenomena is the one referred to by Weinberg in his above-mentioned footnote on Glashow's paper. As noted by Ross and Veltman, "for experimental purposes the difference is that in the Weinberg model the mass of the neutral vector boson is fixed relative to the charged masses if the mixing angle is known." (Ross and Veltman 1975, p. 136). In other words, Weinberg's model predicts a relation between the "mixing angle", which accounts for the relative strengths of the neutral and charged weak interactions and the masses of the Z and W bosons. This relation can be determined experimentally and agrees perfectly with the prediction. With the increase in experimental precision due to large particle colliders like LEP, its importance has even advanced from a test of Weinberg's theory to a probe of quantum effects. For example, the precise measurement of the mixing angle allowed the prediction (assuming that the SM is correct) of the top quark mass before its discovery at the Tevatron. Until recently, it also provided the most stringent constraints on the Higgs boson mass.

Nevertheless, in spite of this highly predictive character, it is questionable whether the relation between the mixing angle and the Z and W boson masses pointed out by Ross and Veltman counts as independent support for the SMHM. What underlies this relation is only the gauge structure of the Higgs-field, not its more specific features as a fundamental scalar which give rise to the worries discussed in Section 3. This prediction of Weinberg's model would be the same for a non-fundamental Higgs-field which transforms under the same gauge representation. The fact that Weinberg chose the minimal representation (i.e. the one that leads to the minimum number of physical Higgs bosons) is hardly ever seriously criticized by physicists (even though models with alternative gauge structures were considered later).

Other aspects that might be proposed as independent experimental support for the SMHM prior to the discovery of the Higgs boson are rejected more easily: for example, the fact that the SMHM provides mass to more than one particle obviously cannot be taken as independent evidence, as each mass requires an additional, arbitrary parameter in the SMHM. CP-violation in the quark sector, as another example, is tied to the existence of non-vanishing quark masses, not to the mechanism that provides these masses. In conclusion, until the recent discovery at the LHC of what could be the Higgs boson, criterion number 1 of the condition of justification was met.

To summarize the analysis of this and the previous section, calling the SMHM "ad hoc" before the discovery of the Higgs boson is very much in the spirit of philosophers' usage of this term. Nevertheless, the SMHM does not, strictly speaking, qualify as "ad hoc" according to Leplin's account (and similar clauses in other philosophers' definitions of "ad hoc hypothesis") due to a conflict with the *condition of experimental anomaly*. We have not proposed an explicit reformulation of this condition, but rather suggested that the conflict could be easily avoided by an appropriate modification, as the essentials of Leplin's definition lie in the other conditions and in the unproblematic core of the condition of experimental anomaly. In the remainder of this paper, we will adopt the "ad hoc" attribute for the SMHM, as long as we refer to the time before July 4, 2012. The situation after the Higgs discovery will be the subject of the next section.

## 7. The status of the SMHM after the discovery of a Higgs-like particle

As a consequence of the recent discovery of a Higgs-like particle, there is now independent experimental evidence for the SMHM: the signal of a particle which the SMHM says should exist. Therefore, the most crucial characteristic of an ad hoc hypothesis, formulated in the first criterion of Leplin's *condition of justification*, is no longer obeyed. According to Leplin's account which we have used as the basis for our analysis, the ad hoc character of the SMHM therefore has ended after the recent discovery of a Higgs-like particle.

It is remarkable that this rather drastic consequence is paralleled by similarly dramatic reactions in the scientific community. Many physicists – and with every further experimental corroboration of the SMHM there will be more of them – now seem to be ready to accept the SMHM as part of physical reality. On the other hand, most of them seem to be *not* ready to conclude that the criticisms of the SMHM were unfounded. Except for the issue of lacking experimental evidence, all the criticisms described in Section 3 still hold. The experimental confirmation of the SMHM does not solve any of these problems. Quite the contrary; it is the end of its ad hoc character which makes them more pressing today.

The discovery of the Higgs boson is too recent to fully assess its overall effects on the physics community. Two possible conclusions that might be drawn by physicists from the discovery of the Higgs boson will be described in the following. The first one is to expect that the experimental confirmation of the SMHM, combined with the problems outlined in Section 3, will be followed by the imminent discovery of "new physics" which will provide a solution to these problems. The latest results[13] do not support this conclusion though; they rather exclude more and more of the proposed alternative theories, or at least severely restrict their allowed parameter space. The search will be intensified once the LHC resumes operation in 2015 at increased center-of-mass energy.

---

13  The latest results can be accessed from the following URLs (retrieved on February 17, 2014) for the ATLAS and CMS experiments, respectively: https://twiki.cern.ch/twiki/bin/view/AtlasPublic/WebHome#Physics_papers; https://twiki.cern.ch/twiki/bin/view/CMSPublic/PhysicsResults

The second possible conclusion to which more and more physicists seem to be willing to subscribe is to critically reconsider the validity (or applicability) of the issues with the SMHM listed in Section 3, first and foremost the unnaturalness criticism. For example, according to theoretical physicist Jonathan Feng:

> For decades, the unnaturalness of the weak scale has been the dominant problem motivating new particle physics […]. This paradigm is now being challenged by a wealth of experimental data. (Feng 2013, p. 1)

In a similar vein, the theoretical particle physicist Michael Krämer writes in a blog entry:

> Naturalness arguments did provide useful insight into particle physics in the past [...] but they do not work all the time. Will the naturalness problem of the Higgs mechanism lead to profound new insights or is it just a red herring? It is still too early to conclude, but the lack of any signal of new physics at the LHC, and the landscape of solutions of string theory,[14] have led us to reconsider the role of naturalness as a particle physics paradigm. (Krämer 2013)

An alternative route to a more relaxed stance towards the naturalness problem would be to re-think its conceptual presuppositions. For example, Christof Wetterich (2012) argues that the need for fine-tuning indicates nothing more than "a shortcoming of the perturbative expansion series" (p. 573) and interprets the measured value of the Higgs mass as an indication that the SM may remain valid up to the Planck scale. On his view, the SMHM is neither more tentative nor less fundamental than the other parts of the SM.

To conclude, the end of the ad hoc character of the SMHM due to the recent discovery, combined with the lack of hints for new physics, apparently leads to a shift in perspective as regards the naturalness problem in particular and the conceptual foundations of the SMHM in general. Of course, the criticisms mentioned in Section 3 are still regarded as serious among physicists, but their relevance and force may now be seen in a different light, since the SMHM has lost its most conspicuous characteristic of ad hocness.

**8. The role of the notion of ad hocness in theory appraisal**

In this final section, we shall confront our main findings in the previous sections with a recent criticism by Hunt that "what is "ad hoc" seems to be a judgment made by particular scientists not on the basis of any well-established definition but rather on their individual aesthetic senses" (Hunt 2012, p. 1). In his view, "a hypothesis considered ad hoc can apparently be retroactively declared non–ad hoc on the basis of subsequent data, rendering the term meaningless" (Ibid.). As we have argued in Section 6, although the ad hoc charge against the SMHM never rested on a "well-established definition", as pointed out by Hunt, it does obey the key characterizations of what philosophers consider an ad hoc hypothesis. As our considerations show, physicists' charges of ad hocness against the SMHM rest on various arguments that are based on widely used, rationally and empirically justifiable, methodological principles of particle physics: for example, that methodologically acceptable instances of symmetry breaking are dynamical rather than fundamental, that the values of fundamental parameters should not involve excessive fine-tunings, etc. At least as far as the SMHM is concerned, it is therefore unjustified to claim that the grounds on which the ad hoc charge rests are "merely aesthetic", safe perhaps in an extremely wide and stretched sense of "aesthetic". Thus, based on our considerations concerning the SMHM, we dispute the claim made by Hunt that scientists' allegations of ad hocness reflect nothing more substantial than their "individual aesthetic sense."

---

14 The idea alluded to here is that considerations involving the string theory landscape may help dispelling the naturalness problem by integrating the SM in a multiverse scenario, where anthropic arguments may be used to explain away the perceived fine tuning of fundamental parameters as an unproblematic observation-selection effect. See (Donoghue 2007) for details.

Arguably, not only the criticisms of the SMHM themselves, but also the label "ad hoc" to frame them go beyond the scientists' "individual aesthetic sense". As we have argued, even though in the light of these criticisms the SMHM is not literally ad hoc according to Leplin's definition, it is certainly ad hoc according to its spirit. More importantly, some of Leplin's criteria of ad hocness capture precisely the point of these criticisms in more general terms.

The considerations of the previous section can be used to shed light on the actual role and significance of the notion of "ad hoc" in practice: according to Leplin's *condition of tentativeness*, part of what constitutes an ad hoc hypothesis is that it may well turn out to be true. The recent discovery of the Higgs boson indicates that this may indeed be the case for the SMHM. With the ensuing end of the ad hoc character of the SMHM (in the sense discussed in the previous section) all the associated problems mentioned in Section 3 – in particular the naturalness problem – have become more acute. Physicists knew that the experimental confirmation of the SMHM without any accompanying hints at new physical phenomena associated with accessible energy scales would shake their confidence in some of their most basic beliefs concerning physical theories, for example that "fine-tuning" of parameters, such as the Higgs mass, is to be avoided; this is why the ad hoc charge has always been taken very seriously.

As a result, contrary to what Hunt suggests about ad hoc hypotheses, the main message we would like to convey at the end of the present paper is neither that the notion of an ad hoc hypothesis is unhelpful nor that the SMHM should never have been called "ad hoc" in the first place. Rather, to conceive of the SMHM as an ad hoc hypothesis—in Leplin's sense—that has finally become supported by independent empirical evidence helps understanding how scientists re-evaluate the hypothesis considered ad hoc in the light of this novel evidence. In particular, the fact that the SMHM seems to be realized in nature despite physicists' strong reservations against it may have a seemingly very paradoxical consequence: that physicists would ultimately have to develop a much more radically new approach to elementary particle physics than if new physics beyond the SM had been discovered, which would complement or even replace the SMHM. In the latter case, the SM might have turned out to be like an effective version of a more elaborate theory, and as put by Ross and Veltman in an already quoted passage, the SMHM might have been replaced with a more sophisticated mechanism for mass generation. Interpreting the experimental observation as showing that the SMHM is *not* merely an "effective description", on the other hand, would imply that physicists' presently preferred criteria of theory choice in elementary particle physics have serious shortcomings. This could require a much more profound change of ideas concerning particle physics than the discovery of "new physics" (such as supersymmetry, for example). From this perspective, the physicists' classification of the SMHM as ad hoc may have verbalized their resistance against such profound change.

**Acknowledgments** This research is part of the project "An Ontological and Epistemological Analysis of the Higgs-mechanism," funded by the Deutsche Forschungsgemeinschaft (DFG, contract HA 2990/4-1), within the research collaboration "The Epistemology of the Large Hadron Collider (LHC)" at the University of Wuppertal: http://www.lhc-epistemologie.uni-wuppertal.de. The authors would like to thank two anonymous referees, as well as audiences at conferences and seminars in Ankara, Dresden, Tel Aviv and Wuppertal, for thoughtful comments and suggestions.